\documentclass[12pt]{article}
\usepackage{amssymb}
\usepackage{amsmath}
\usepackage[dvips]{graphicx}

\usepackage[dvips]{epsfig}

\begin{document}

\author{A. de Souza Dutra$^{a,b}$\thanks{%
dutra@feg.unesp.br} and R. A. C. Correa$^{b}$\thanks{
fis04132@feg.unesp.br} \\
\\
$^{a}$Abdus Salam ICTP, Strada Costiera 11, Trieste, I-34100 Italy.\\
$^{b}$UNESP-Campus de Guaratinguet\'{a}-DFQ\thanks{
Permanent address}\\
Departamento de F\'{\i}sica e Qu\'{\i}mica\\
12516-410 Guaratinguet\'{a} SP Brasil}
\title{\textbf{Fluctuating solutions for the evolution of domain walls}}
\maketitle

\begin{abstract}
A class of oscillating Lorentz covariant configurations for the evolution of
the domain walls in diverse dimensions are analytically obtained. It is
shown that the oscillating solutions in the case of domain walls are
responsible for structures which are larger than the usual kink-like
configurations and, in the Lorentz covariant evolution case, lead to
long-standing configurations.
\end{abstract}

\newpage

\section{Introduction}

The study of nonlinear systems have been growing since the sixties of the
last century \cite{b1,b2}. Nowadays the nonlinearity is found in many areas
of the physics, including condensed matter physics, field theory, cosmology
and others \cite{rajaramanweinberg}-\cite{b6.1}. Particularly. whenever we
have a potential with two or more degenerate minima, one can find different
vacua at different portions of the space. Thus, one can find domain walls
connecting such regions.

In a beautiful and seminal work, Coleman \cite{coleman,coleman2} described
what it was called the \textquotedblleft fate of the false
vacuum\textquotedblright\ through a semiclassical analysis of an asymmetric $%
\lambda \phi ^{4}$-like model. In such situation the decaying process of the
field configuration from the local to the global vacuum of the model is
analyzed. In a very recent work by Dunne and Wang \cite{dunne}, it was shown
that in the presence of gravity some fluctuating bounce solutions show up.
In that work, those oscillating bounces where numerically obtained. Here we
will show that those beautiful unusual field configurations appear already
in the case where there are no gravitational interaction. Interestingly, in
the work of Dunne and Wang, it has been asserted that those fluctuating
solutions do not appear in the flat-space limit. However, in their work they
analyzed the system through an Euclidean action. Here, instead, we work with
a Minkowski space-time metric. Furthermore, they were concerned with
instanton solutions ($\phi \left( t\right) $), and we deal with kinks and
lumps ($\phi \left( x\right) $).

As observed by Coleman \cite{coleman}, this kind of system can be used to
describe nucleation processes on statistical physics, crystallization of a
supersaturated solution, the boiling of a superheated fluid and even in the
case of the evolution of cosmological models. \ In this last application,
one can suppose that when the universe have been created it was far from any
vacuum state. As it has expanded and cooled down it evolved first to a false
vacuum instead of the true one. Thus, in such a scenario, when the time goes
by, the universe should finally be settled in the true vacuum state. As we
are going to see below, an evident and important consequence of the
existence of these oscillating solutions is that the decaying process can be
retarded when compared with the non-fluctuating configurations. Furthermore,
if one takes into account the case of static domain walls, it is clear that
the oscillating configurations are responsible for structures which are
larger than the usual kink-like configurations. So one can use those
solutions in order to describe thicker walls.

Besides, as we will see below, there are some covariant
fluctuating configurations, which are capable to describe systems
where the configuration evolves from the false to the true vacuum.
The duration of the evolution from a vacuum to another, depends on
the number of oscillations. As observed above, this can be used to
describe some possible long-standing cosmological evolutions.
Moreover, one can think also about ferromagnetic systems which,
beginning with randomly distributed magnetic domains, are
submitted to an external magnetic field. In this situation, the
domains which present an orientation of their magnetization along
the direction of the external field, can be thought as living in
the true vacuum. The remaining will be in the false ones and, as a
consequence, will tend to evolve to the true vacuum by aligning
themselves with the external magnetic field. This process probably
will not happens suddenly but, on the contrary, they might
oscillate before reaching the true vacuum. This process would take
a longer time to occur, when compared with a direct evolution from
the false to the true vacuum, as we see below in this work.

Aiming to achieve those analytical solutions, we are concerned
with a model which is a modification of the one introduced by
Horovitz, following a suggestion of S. Aubry, when studying the
presence of solitons in discrete chains, Horovitz and
collaborators \cite{horovitz} introduced the so-called
Double-Quadratic (DQ) \cite{deleo}-\cite{losano} model, whose
potential is given by
\begin{equation}
V(\phi )=\frac{1}{2}\phi ^{2}-\left\vert \phi \right\vert +\frac{1}{2}.
\end{equation}

\begin{figure*}[tbp]
\centering
\includegraphics{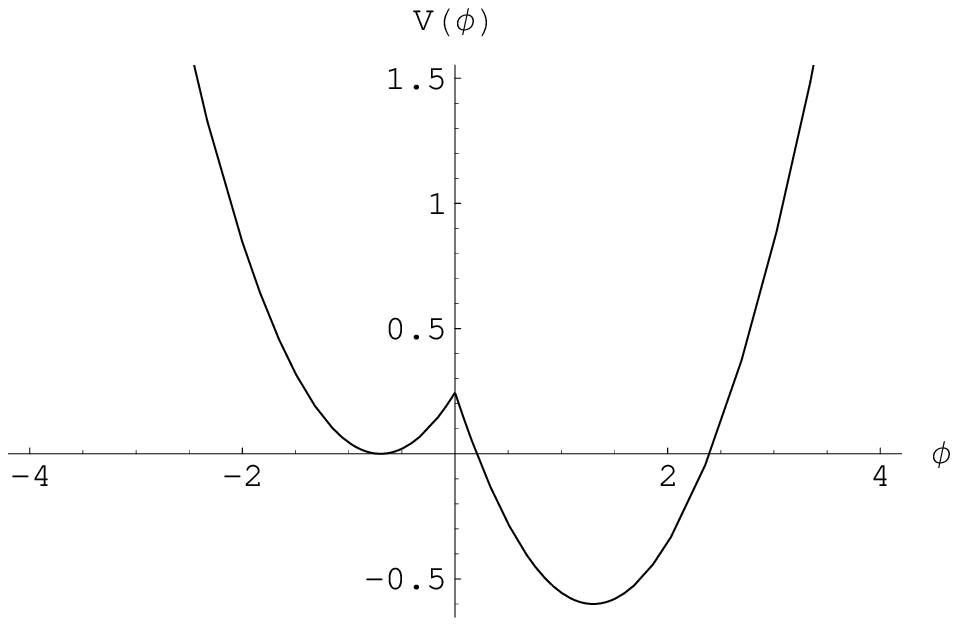}
\caption{Potential of the asymmetrical doubly-quadratic (ADQ) model for $%
\protect\epsilon =0.3$.}
\label{fig1:potencialmodelo}
\end{figure*}

That mentioned model was introduced by Stavros Theodorakis \cite{b6}. It
allows one to obtain explicit analytical solutions for the kink-like and
lump-like field configurations, these last were called in that paper as
critical bubbles. The model introduced in \cite{b6} is an asymmetrical
version of the DQ model (ADQ), and it is characterized by the potential

\begin{equation}
V(\phi )=\frac{1}{2}\phi ^{2}-\left\vert \phi \right\vert -\epsilon \phi +%
\frac{1}{2}(\epsilon -1)^{2},  \label{1}
\end{equation}%
\noindent where $0<\epsilon <1$. In Fig. \ref{fig1:potencialmodelo} this
potential is plotted. In fact, a similar potential was used in order to
analyze the case of wetting and oil-water-surfactant mixtures \cite{gompper}%
. So, the solutions described here will may also have important impact over
this matter.

It can be seen that it is a kind of asymmetrical $\lambda \phi
^{4}$ model, where the potential has a derivative discontinuity at
$\phi =0$, what was called a \textquotedblleft kink" in \cite{b6},
and this is in contrast with the usual meaning used for the word
kink in the quantum field theory literature. Here we use this last
nomenclature instead, where the word kink stands for a field
configuration which interpolates between different vacua of the
model, and lump corresponds to a configuration where the field
interpolates a vacuum with itself. In the next sections, we will
construct explicitly some examples of the new class above
mentioned, and discuss the physical features of the analytical
solutions. Finally, we will present a generalization for an
arbitrary number of oscillations at the conclusions section. In
particular, in Section 4, the extension of the evolution domain
walls \cite{b6} is taken into account.

\section{New kinks and lumps in one spatial dimension}

In this section we present the first example of an entire new class of
static lump and kink-like solutions for the ADQ model. For this, we suppose
that there are four regions where the field presents alternating signals.
Specifically, we look for a solution where $\phi _{I}<0$, $\phi _{II}>0$, $%
\phi _{III}<0$ and $\phi _{IV}>0$. As a consequence, the points, $x_{1}$, $%
x_{2}$ and $x_{3}$ which separate these regions, shall correspond to zeros
of $\phi (x)$. \

The finiteness of the solution along the whole spatial axis imply into a
restriction of the integration constants. Additionally, we require that the
field approaches asymptotically the vacua of the model, respectively given
by $\phi _{I}(-\infty )=-1+\epsilon $ and $\phi _{IV}(\infty )=1+\epsilon $.
Moreover, since the field must vanish at the intermediate points, we shall
grant that $\phi _{I}(x_{1})=\phi _{II}(x_{1})=0$, $\phi _{II}(x_{2})=\phi
_{III}(x_{2})=0$ e $\phi _{IV}(x_{3})=\phi _{IV}(x_{3})=0$. Thus, we finish
with

\begin{eqnarray}
\phi _{I}(x) &=&(1-\epsilon )(e^{x-x_{1}}-1),  \notag \\
\phi _{II}(x) &=&(1+\epsilon )\left[ 1-\frac{\sinh (x-x_{1})}{\sinh
(x_{2}-x_{1})}-\frac{\sinh (x_{2}-x)}{\sinh (x_{2}-x_{1})}\right] ,  \notag
\\
&& \\
\phi _{III}(x) &=&-(1-\epsilon )\left[ 1-\frac{\sinh (x-x_{2})}{\sinh
(x_{3}-x_{2})}-\frac{\sinh (x_{3}-x)}{\sinh (x_{3}-x_{2})}\right] ,  \notag
\\
\phi _{IV}(x) &=&(1+\epsilon )(1-e^{x_{3}-x}).  \notag
\end{eqnarray}

\begin{figure*}[tbp]
\centering
\includegraphics{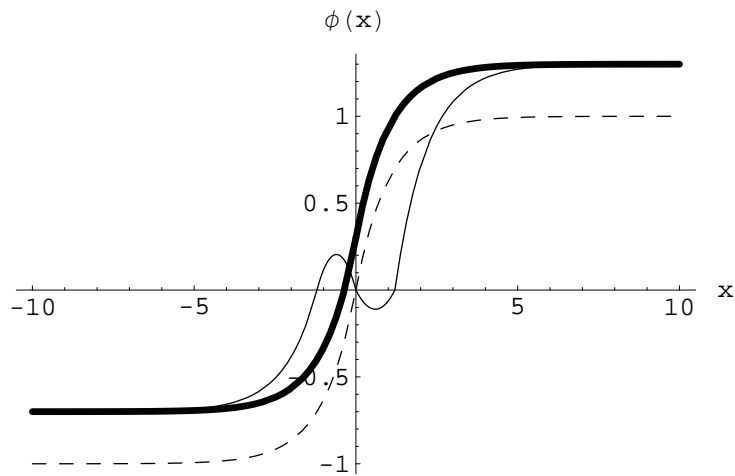}
\caption{Kinklike solutions. The dashed line corresponds to the
solution presented by Theodorakis for the case where
$\protect\epsilon =0$. The thick line is the profile of that
configuration in the asymmetrical case, and the thin line is the
new kink (Eq. (3)), both plotted using $\protect\epsilon =0.3$. }
\label{fig2:solkinks}
\end{figure*}

The above kink-like configuration is represented in Fig. \ref{fig2:solkinks}%
. Note however that, in order to be sure that the first derivative of the
field configuration is continuous at the intermediate point, one must
constrain the distance between those points when $0<\epsilon <1$ through the
relation

\begin{equation}
\frac{1-\cosh (x_{2}-x_{1})}{\sinh (x_{2}-x_{1})}=\frac{\epsilon -1}{%
\epsilon +1}=\frac{1-\cosh (x_{3}-x_{2})}{\sinh (x_{3}-x_{2})},  \label{30}
\end{equation}%
which, after some algebraic manipulations, is equivalent to require that

\begin{equation}
x_{2}-x_{1}=x_{3}-x_{2}=ArcCoth\left( \frac{1+\epsilon ^{2}}{1-\epsilon ^{2}}%
\right) .  \label{30.1}
\end{equation}

Here, it is important to remark that the case where $\epsilon =0$ can not be
recovered from the one with $0<\epsilon <1$. In this particular situation,
one can see that the differences $x_{2}-x_{1}$ and $x_{3}-x_{2}$ become
equal to zero and we reproduce the case considered in \cite{b6}.

Now, performing the analysis of the linear stability of this new
solution by using the usual procedure \cite{Rajaraman}, one ends
with the following transcendental equation

\begin{eqnarray}
&&\frac{\alpha }{2}\left[ 1-\frac{1}{\omega (1-\epsilon )}\right] e^{2\omega
x_{2}}-\frac{1}{\omega (1-\epsilon )}\left[ 1-\frac{\alpha }{2}\right] +
\notag \\
&&\left. -\left( \frac{2+\beta }{\beta }\right) \left\{ \frac{\left( \alpha
+2\right) }{2}\left[ 1-\frac{1}{\omega (1-\epsilon )}\right] +\frac{\alpha
e^{-2\omega x_{3}}}{2\omega (1-\epsilon )}\right\} e^{\omega x_{3}}=0,\right.
\label{30.2}
\end{eqnarray}

\noindent with $\omega =\sqrt{1-E}$, $\alpha \equiv -2/(\omega
\left\vert d\phi _{2}/dx\right\vert _{x=x_{2}})$ and $\beta \equiv
-2/(\omega \left\vert d\phi _{3}/dx\right\vert _{x=x_{3}})$. For
instance, by analyzing an example where $\epsilon =0.367$, one can
verify that there are two solutions with negative energies
$E=-10.7113$ and $E=-1.49798$. This implies that the solution is
unstable.

Once again, it was possible to find a kink-like solution, but this time
presenting an oscillating region which can become larger if one introduces
more regions in the game, as we are going to see later.

Since the model presents these two vacua, one can ask for other analytical
solutions connecting a given vacuum to itself. In fact, one can construct
two of these solutions, one for the false vacuum and another for the true
one.

\begin{figure*}[tbp]
\centering
\includegraphics{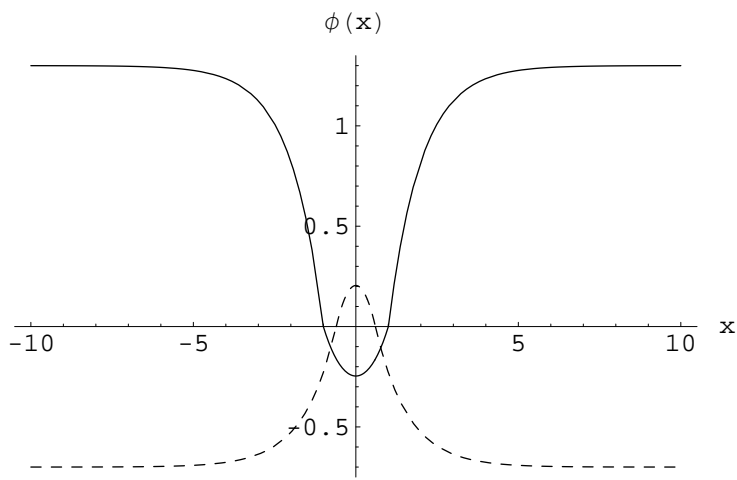}
\caption{Lumplike solutions (bubble) in one dimension for $\protect\epsilon %
=0.3$. The dashed line is the solution presented by Theodorakis for a
configuration starting and finishing at the false vacuum and the solid line
is our solution connecting the true vacuum to itself.}
\label{fig6:bolhasimplesnossaStavros1D}
\end{figure*}

Below, we will determine a solution which presents three regions, beginning
and ending at the true vacuum. In certain sense this solution completes the
simplest set of lumps or critical bubbles which were introduced by
Theodorakis in \cite{b6} (starting and finishing at the false vacuum). In
Fig. \ref{fig6:bolhasimplesnossaStavros1D}, it can be seen a plot of both
ours and the Theodorakis solutions.

Now, the field $\phi $ in each one of the regions will behave respectively
as $\phi _{I}>0$, $\phi _{II}<0$ e $\phi _{III}>0$. In this case, we can
show that it is a stable solution, in contrast with the one analyzed by
Theodorakis. In other words, the lump connecting the true vacuum to itself
is stable and the one related to the false vacuum is not. Thus, by making an
analysis similar to that one done by Theodorakis, we can interpret this
result in the sense that there is a stable portion of false vacuum among two
regions of true vacua.

Furthermore, we are interested in constructing more complex solutions, where
the size of the region separating the true and the false vacua is larger.
For this, we introduce an entire new class of solutions, where the field
oscillates between the true and the false vacua regions. Let us present an
example with five regions such that $\phi _{I}<0$, $\phi _{II}>0$, $\phi
_{III}<0$, $\phi _{IV}>0$ and $\phi _{V}<0$. Again, the points $x_{1}$, $%
x_{2}$, $x_{3}$ and $x_{4}$ correspond to zeros of $\phi (x)$.

With this in mind, it is not difficult to conclude that \ one should have
the following solution

\begin{eqnarray}
\phi _{I}(x) &=&(1-\epsilon )(e^{x-x_{1}}-1),  \notag \\
\phi _{II}(x) &=&(1+\epsilon )\left[ 1-\frac{\sinh (x-x_{1})}{\sinh
(x_{2}-x_{1})}-\frac{\sinh (x_{2}-x)}{\sinh (x_{2}-x_{1})}\right] ,  \notag
\\
\phi _{III}(x) &=&-(1-\epsilon )\left[ 1-\frac{\sinh (x-x_{2})}{\sinh
(x_{3}-x_{2})}-\frac{\sinh (x_{3}-x)}{\sinh (x_{3}-x_{2})}\right] ,
\label{38} \\
\phi _{IV}(x) &=&(1+\epsilon )\left[ 1-\frac{\sinh (x-x_{3})}{\sinh
(x_{4}-x_{3})}-\frac{\sinh (x_{4}-x)}{\sinh (x_{4}-x_{3})}\right]   \notag \\
\phi _{V}(x) &=&(\epsilon -1)(1-e^{x_{4}-x}).  \notag
\end{eqnarray}

This is the case in which we apply our idea to the case explored by
Theodorakis, when he was considering the critical bubble in the false
vacuum. The difference is that, now, the field configuration starting at the
false vacuum, before going back to it, oscillates once between the positive
and negative vacua regions.

Verifying the stability of this new solution, one can see that the potential
$V_{st}=1-2~\delta \lbrack \phi (x)]$, where
\begin{equation}
\delta \lbrack \phi (x)]=\frac{\delta (x-x_{1})}{\left\vert \frac{d\phi }{dx}%
\right\vert _{x=x_{1}}}+\frac{\delta (x-x_{2})}{\left\vert \frac{d\phi }{dx}%
\right\vert _{x=x_{2}}}+\frac{\delta (x-x_{3})}{\left\vert \frac{d\phi }{dx}%
\right\vert _{x=x_{3}}}+\frac{\delta (x-x_{4})}{\left\vert \frac{d\phi }{dx}%
\right\vert _{x=x_{4}}}.  \label{40.1}
\end{equation}%
This time the equation used to find the energies of the bound states related
to the perturbation is expressed as

\begin{eqnarray}
&&\left\{ \beta \lbrack 1-(1+\gamma )]e^{2\omega x_{4}}-\gamma (1-\beta
)\right\} \left\{ \left( \frac{\alpha }{2}-1\right) \left[ \frac{1}{\omega
(\epsilon -1)}\right] \right.  \notag \\
&&\left. -\frac{\alpha }{2}\left[ 1+\frac{1}{\omega (\epsilon -1)}\right]
e^{2\omega x_{2}}\right\} -\{(\beta +1)(\gamma +2)e^{2\gamma \omega
x_{4}}-\beta \gamma e^{2\gamma \omega x_{3}}\}  \notag \\
&&\left. \times \left\{ 1-\left( \frac{\alpha }{2}-1\right) \left[ \frac{1}{%
\omega (\epsilon -1)}\right] +\frac{\alpha }{2}\left[ 1+\frac{1}{\omega
(\epsilon -1)}\right] e^{2\omega x_{2}}\right\} =0\right.  \label{40.2}
\end{eqnarray}

\noindent with

\begin{equation*}
\alpha \equiv -\frac{2}{\omega \left\vert \frac{d\phi _{2}}{dx}\right\vert
_{x=x_{2}}},~\beta \equiv -\frac{2}{\omega \left\vert \frac{d\phi _{3}}{dx}%
\right\vert _{x=x_{3}}}\text{ and }\gamma \equiv -\frac{2}{\omega \left\vert
\frac{d\phi _{4}}{dx}\right\vert _{x=x_{4}}}.
\end{equation*}%
From the equation (\ref{40.2}) we get a negative value for the energy,
signalizing that this configuration, as the one in \cite{b6}, is unstable.

The corresponding configuration starting and finishing at the true vacuum is
given by

\begin{align}
\phi _{I}(x)& =(1+\epsilon )(1-e^{x-x_{1}}),  \notag \\
\phi _{II}(x)& =-(1-\epsilon )\left[ 1-\frac{\sinh (x-x_{1})}{\sinh
(x_{2}-x_{1})}-\frac{\sinh (x_{2}-x)}{\sinh (x_{2}-x_{1})}\right] ,  \notag
\\
\phi _{III}(x)& =(1+\epsilon )\left[ 1-\frac{\sinh (x-x_{2})}{\sinh
(x_{3}-x_{2})}-\frac{\sinh (x_{3}-x)}{\sinh (x_{3}-x_{2})}\right] ,
\label{40.5} \\
\phi _{IV}(x)& =-(1-\epsilon )\left[ 1-\frac{\sinh (x-x_{3})}{\sinh
(x_{4}-x_{3})}-\frac{\sinh (x_{4}-x)}{\sinh (x_{4}-x_{3})}\right]   \notag \\
\phi _{V}(x)& =(1+\epsilon )(1-e^{x_{4}-x}).  \notag
\end{align}%
These last two solutions are plotted in Fig. \ref%
{fig7:bolhanossa1dfalsetruevacuum}. Again, it is not difficult to verify
that it is unstable.

\section{Fluctuating solitons in two and three dimensions}

In two spatial dimensions, the spherically symmetric static configurations
of the ADQ model must obey the equation

\begin{equation}
\frac{\partial ^{2}\phi }{\partial r^{2}}+\frac{1}{r}\frac{\partial \phi }{%
\partial r}=\phi -\frac{\phi }{\left\vert \phi \right\vert }-\epsilon ,
\label{41}
\end{equation}

\noindent where $r=\sqrt{x^{2}+y^{2}}$. Thus, when $r\longrightarrow \infty $
the field $\phi \longrightarrow \epsilon -1$.

Now, we construct a solution with four distinct solutions. These regions
are, once again, characterized by $0\leq \phi _{1}\leq r_{1}$, $r_{1}\leq
\phi _{2}\leq r_{2}$, $r_{2}\leq \phi _{3}\leq r_{3}$, $r_{3}\leq \phi _{4}$%
, with $\phi _{1}>0$, $\phi _{2}<0$, $\phi _{3}>0$ e $\phi _{4}<0$. After
imposing the continuity of the field in each transition point, one gets

\begin{eqnarray}
\phi _{1}(r) &=&(1+\epsilon )\left[ 1-\frac{I_{0}(r)}{I_{0}(r_{1})}\right] ,
\notag \\
\phi _{2}(r) &=&(1-\epsilon )\left\{ \frac{%
[K_{0}(r_{1})-K_{0}(r_{2})]I_{0}(r)+[I_{0}(r_{2})-I_{0}(r_{1})]K_{0}(r)}{%
K_{0}(r_{1})I_{0}(r_{2})-K_{0}(r_{2})I_{0}(r_{1})}+\right.   \notag \\
&&\left. -1\right\} , \\
\phi _{3}(r) &=&(1+\epsilon )\left\{ \frac{%
[K_{0}(r_{2})-K_{0}(r_{3})]I_{0}(r)+[I_{0}(r_{3})-I_{0}(r_{2})]K_{0}(r)}{%
K_{0}(r_{3})I_{0}(r_{2})-K_{0}(r_{2})I_{0}(r_{3})}+\right. ,  \notag \\
&&\left. +1\right\} ,  \notag \\
\phi _{4}(r) &=&(\epsilon -1)\left[ 1-\frac{K_{0}(r)}{K_{0}(r_{3})}\right] .
\notag
\end{eqnarray}

The functions $K_{0}(r)$ and $I_{0}(r)$ are the modified Bessel functions.
This oscillating kink solution is represented in Fig. \ref%
{fig10:bol2d}

The stability equation in this dimension is given by

\begin{equation}
\left( -\frac{d^{2}}{dr^{2}}-\frac{1}{r}\frac{d}{dr}+V_{st.}\right) \Theta
(r)=E~\Theta (r),  \label{49.1}
\end{equation}

\noindent where we have $V_{st}=1-2~\delta \lbrack \phi (r)]$.

By solving Eq. (\ref{49.1}), we find a positive value for the ground state,
indicating that this configuration in two dimensions is stable, so that, for
a small initial perturbation the solution is kept finite when the time goes
by. In fact, the value of the ground state solution of (\ref{49.1}) found
for this configuration is $E_{0}=0.324$ ($\epsilon =0.3$). It is important
to remark that, in this case, the Theodorakis solution is also stable,
having $E=~0.592$.

Let us end this section by taking into account the three-dimensional case.
In this dimension the spherically symmetric static field configurations must
obey the equation

\begin{equation}
\frac{\partial ^{2}\phi }{\partial r^{2}}+\frac{2}{r}\frac{\partial \phi }{%
\partial r}=\phi -\frac{\phi }{\left\vert \phi \right\vert }-\epsilon ,
\label{50}
\end{equation}

\noindent where $r=\sqrt{x^{2}+y^{2}+z^{2}}$. Below we construct a solution
which presents the same features of the one discussed in the previous case.
After applying the continuity conditions, one can conclude that

\begin{eqnarray}
\phi _{1}(r) &=&(1+\epsilon )\left[ 1-\frac{r_{1}}{r}\frac{\sinh (r)}{\sinh
(r_{1})}\right] ,  \notag \\
\phi _{2}(r) &=&(\epsilon -1)\left[ \frac{(r_{1}e^{r_{1}}-r_{2}e^{r_{2}})}{%
(e^{2r_{2}}-e^{2r_{1}})}\frac{e^{r}}{r}+\frac{(r_{2}e^{r_{1}}-r_{1}e^{r_{2}})%
}{2re^{r}} {csch}(r_{2}-r_{1})+\right.   \notag \\
&&\left. +1\right] , \\
\phi _{3}(r) &=&(\epsilon +1)\left[ \frac{(r_{3}e^{r_{3}}-r_{2}e^{r_{2}})}{%
(e^{2r_{2}}-e^{2r_{3}})}\frac{e^{r}}{r}+\frac{(r_{3}e^{r_{2}}-r_{2}e^{r_{3}})%
}{2re^{r}} {csch}(r_{3}-r_{2})+\right.   \notag \\
&&\left. +1\right] ,  \notag \\
\phi _{4}(r) &=&(1-\epsilon )\left[ \frac{r_{3}}{r}e^{r_{3}-r}-1\right] .
\notag
\end{eqnarray}

In order to determine the stability of the configuration, the
equation to be solved in three spatial dimensions is written as

\begin{equation}
\left( -\frac{d^{2}}{dr^{2}}-\frac{2}{r}\frac{d}{dr}+V_{st.}\right) \Theta
(r)=E\Theta (r),  \label{58.1}
\end{equation}

\noindent and, again, $V_{st}=1-2~\delta \lbrack \phi (r)]$. In this case,
we begin by obtaining the solution at each side of the transition points of
the following equation

\begin{equation}
\frac{d^{2}\Theta }{dr^{2}}+\frac{2}{r}\frac{d\Theta }{dr}+(E-1)\Theta =0
\label{58.2}
\end{equation}

\noindent so obtaining

\begin{eqnarray}
\Theta _{1}(r) &=&C_{1}~r^{-1}\sin (r\sqrt{1-E}),  \notag \\
\Theta _{2}(r) &=&C_{2}~r^{-1}\sin (r\sqrt{1-E})+D_{2}~r^{-1}\cos (r\sqrt{1-E%
}),  \notag \\
&& \\
\Theta _{3}(r) &=&C_{3}~r^{-1}\sin (r\sqrt{1-E})+D_{3}~r^{-1}\cos (r\sqrt{1-E%
}),  \notag \\
\Theta _{5}(r) &=&C_{5}~r^{-1}\sin (r\sqrt{1-E}).  \notag
\end{eqnarray}

On the other hand, the discontinuity at the transition points due to the
presence of the Dirac delta function leads us to impose that

\begin{eqnarray}
\left. r\frac{d\Theta }{dr}\right\vert _{r=r_{1}+\xi }-\left. r\frac{d\Theta
}{dr}\right\vert _{r=r_{1}-\xi } &=&\frac{2r_{1}\Theta _{1}(r_{1})}{%
\left\vert \frac{d\phi _{1}}{dr}\right\vert _{r=r_{1}}},  \notag \\
\left. r\frac{d\Theta }{dr}\right\vert _{r=r_{2}+\xi }-\left. r\frac{d\Theta
}{dr}\right\vert _{r=r_{2}-\xi } &=&\frac{2r_{1}\Theta _{2}(r_{2})}{%
\left\vert \frac{d\phi _{2}}{dr}\right\vert _{r=r_{2}}},  \label{58.8} \\
\left. r\frac{d\Theta }{dr}\right\vert _{r=r_{3}+\xi }-\left. r\frac{d\Theta
}{dr}\right\vert _{r=r_{3}-\xi } &=&\frac{2r_{1}\Theta _{3}(r_{3})}{%
\left\vert \frac{d\phi _{3}}{dr}\right\vert _{r=r_{3}}},  \notag
\end{eqnarray}

\noindent where we have taken $\xi \longrightarrow 0$.

Using the continuity at each one of the transition points for the
functions (18) and also the above three conditions, we finish with
a transcendental equation defining a negative energy solution,
showing that this configuration is, once more, unstable.

\section{Lorentz invariant solutions for the oscillating configurations}

Since the Lagrangian density we are dealing with is a Lorentz invariant one,
we can look for configurations for the field $\phi $ depending on an
invariant quantity like $x_{\mu }x^{\mu }=t^{2}-r^{2}$, where $%
r^{2}=\sum\limits_{i=1}^{N}x_{i}^{2}$, and $N$ stands for the number of
spatial dimensions.

Following the work by Theodorakis \cite{b6}, we look for a solution which
describes a transition from the false to the true vacuum. As observed in the
Introduction section, it can be related to a number of interesting physical
systems. In this situation, the scalar field evolves to the value $\epsilon
+1$ when $x_{\mu }x^{\mu }~\rightarrow \infty $ because, when this happens,
the field approaches its true vacuum. Besides, it will be equal to $\epsilon
-1$ if $x_{\mu }x^{\mu }~\rightarrow -~\infty $. As we are interested in the
construction of a new family of configurations, we propose a scalar field $%
\phi $ with four regions, where $\phi _{I}<0$, $\phi _{II}>0$, $\phi _{III}<0
$ and $\phi _{IV}>0,$ similarly to the case in which we have introduced the
oscillating kink-like configuration.

First of all, we define the variable $s=t^{2}-r^{2}\equiv x_{\mu }x^{\mu }$.
Thus, we will have regions where $t^{2}-r^{2}>0$ and $t^{2}-r^{2}<0$. For
the case where $t^{2}-r^{2}>0$, the field equation is written as

\begin{equation}
\frac{\partial ^{2}\phi }{\partial \chi ^{2}}+\frac{N}{\chi }\frac{\partial
\phi }{\partial \chi }+\phi -\frac{\phi }{\left\vert \phi \right\vert }%
-\epsilon =0,  \label{1.11}
\end{equation}

\noindent where $\chi =\sqrt{t^{2}-r^{2}}=\sqrt{s}$. Otherwise, if $%
t^{2}-r^{2}<0$ we get

\begin{equation}
\frac{\partial ^{2}\phi }{\partial \rho ^{2}}+\frac{N}{\rho }\frac{\partial
\phi }{\partial \rho }-\phi +\frac{\phi }{\left\vert \phi \right\vert }%
+\epsilon =0,  \label{1.12}
\end{equation}

\noindent with $\rho =\sqrt{r^{2}-t^{2}}=\sqrt{-s}$.

The equations (\ref{1.11}) and (\ref{1.12}) lead us to the solutions

\begin{eqnarray}
\phi (\chi ) &=&\chi ^{-p}[a_{1}J_{p}(\chi )+a_{2}Y_{p}(\chi )]+1+\epsilon ,%
\text{ if }\phi >0\text{, }s^{2}>0  \notag \\
\phi (\chi ) &=&\chi ^{-p}[b_{1}J_{p}(\chi )+b_{2}Y_{p}(\chi )]-1+\epsilon ,%
\text{ if }\phi <0\text{, }s^{2}>0  \notag \\
&& \\
\phi (\rho ) &=&\rho ^{-p}[c_{1}K_{p}(\rho )+c_{2}I_{p}(\rho )]+1+\epsilon ,%
\text{ if }\phi >0\text{, }s^{2}<0  \notag \\
\phi (\rho ) &=&\rho ^{-p}[d_{1}K_{p}(\rho )+d_{2}I_{p}(\rho )]-1+\epsilon ,%
\text{ if }\phi <0\text{, }s^{2}<0  \notag
\end{eqnarray}

\noindent where $p\equiv(N+1)/2$.

After the analysis of the impact of the asymptotic conditions $\phi
\longrightarrow \epsilon \pm 1$ when $x_{\mu }x^{\mu }~\rightarrow \pm
~\infty $, and also considering that $\phi $ must be finite throughout the
range of the validity of the variable $s$, we conclude that the roots of $%
\phi $ shall assume negative values, and we finish with

\begin{eqnarray}
\phi \left( \rho \right)  &=&\epsilon -1+d_{1}\rho ^{-p}~K_{p}(\rho ),\text{
}t^{2}-r^{2}<s_{1}  \notag \\
\phi (\rho ) &=&\rho ^{-p}[c_{1}~K_{p}(\rho )+c_{2}~I_{p}(\rho )]+1+\epsilon
,\text{ }s_{1}<\text{ }t^{2}-r^{2}<s_{2}  \notag \\
\phi (\rho ) &=&\rho ^{-p}[d_{3}~K_{p}(\rho )+d_{4}~I_{p}(\rho )]-1+\epsilon
,\text{ }s_{2}<\text{ }t^{2}-r^{2}<s_{3}  \label{1.19} \\
\phi (\rho ) &=&1+\epsilon +c_{3}~\rho ^{-p}I_{p}(\rho ),\text{ }s_{3}<\text{
}t^{2}-r^{2}<0  \notag \\
\phi (\chi ) &=&1+\epsilon +a_{1}~\chi ^{-p}J_{p}(\chi ),\text{ }%
t^{2}-r^{2}>0.  \notag
\end{eqnarray}

Now, imposing that at each transition point $\phi $ has a root and at same
time granting the continuity at $s=0$, one gets

\begin{eqnarray}
\phi \left( \rho \right)  &=&(\epsilon -1)\left[ 1-\rho ^{-p}(\rho _{1})^{p}%
\frac{K_{p}(\rho )}{K_{p}(\rho _{1})}\right] ,\text{ }t^{2}-r^{2}<s_{1}
\notag \\
\phi (\rho ) &=&(\epsilon +1)\left\{ \rho ^{-p}\left[ -\left( \rho
_{1}^{p}+I_{p}(\rho _{1})\frac{\rho _{1}^{p}K_{p}(\rho _{2})-\rho
_{2}^{p}K_{p}(\rho _{1})}{I_{p}(\rho _{2})K_{p}(\rho _{1})-I_{p}(\rho
_{1})K_{p}(\rho _{2})}\right) \frac{K_{p}(\rho )}{K_{p}(\rho _{1})}\right.
\right.   \notag \\
&&\left. \left. \left. +\left( \frac{\rho _{1}^{p}K_{p}(\rho _{2})-\rho
_{2}^{p}K_{p}(\rho _{1})}{I_{p}(\rho _{2})K_{p}(\rho _{1})-I_{p}(\rho
_{1})K_{p}(\rho _{2})}\right) I_{p}(\rho )\right] +1\right\} ,s_{1}<\text{ }%
t^{2}-r^{2}<s_{2}\right.   \notag \\
\phi (\rho ) &=&(\epsilon -1)\left\{ \rho ^{-p}\left[ \left( \rho
_{2}^{p}+I_{p}(\rho _{2})\frac{\rho _{3}^{p}K_{p}(\rho _{2})-\rho
_{2}^{p}K_{p}(\rho _{3})}{I_{p}(\rho _{3})K_{p}(\rho _{2})-I_{p}(\rho
_{2})K_{p}(\rho _{3})}\right) \frac{K_{p}(\rho )}{K_{p}(\rho _{2})}\right.
\right.  \\
&&\left. \left. \left. +\left( \frac{\rho _{3}^{p}K_{p}(\rho _{2})-\rho
_{2}^{p}K_{p}(\rho _{3})}{I_{p}(\rho _{3})K_{p}(\rho _{2})-I_{p}(\rho
_{2})K_{p}(\rho _{3})}\right) I_{p}(\rho )\right] -1\right\} ,s_{2}<\text{ }%
t^{2}-r^{2}<s_{3}\right.   \notag \\
\phi \left( \rho \right)  &=&(\epsilon +1)\left\{ 1-\rho ^{-p}(\rho _{3})^{p}%
\frac{I_{p}(\rho )}{I_{p}(\rho _{3})}\right\} ,\text{ }s_{3}<\text{ }%
t^{2}-r^{2}<0  \notag \\
\phi (\chi ) &=&(\epsilon +1)\left\{ 1-\chi ^{-p}(\rho _{3})^{p}\frac{%
J_{p}(\chi )}{I_{p}(\rho _{3})}\right\} ,\text{ }t^{2}-r^{2}>0,  \notag
\end{eqnarray}%
with $\rho _{1}=\sqrt{-s_{1}}$, $\rho _{2}=\sqrt{-s_{2}}$ and $\rho _{3}=%
\sqrt{-s_{3}}$. The points $\rho _{1},\rho _{2}$ and $\rho _{3}$
were found under the condition that at each transition between the
regions the scalar field and its first derivative must be
continuous. In Fig. 7, we present a comparison between the
solution appearing in \cite{b6} and the one introduced in this
work.

\section{Conclusions}

In summary, in this work we were able to construct a large class
of new static configurations and, particularly, these solutions
are such that the size of the wall between two regions of a given
vacuum depends on the number of oscillations of the field.
Furthermore, we also studied the stability of each of the
solutions, observing that almost all of them are unstable, except
by the two-dimensional kinks (bubbles in the Theodorakis
nomenclature), and the lump connecting the true vacuum to itself
in one spatial dimension. This lead us to conclude that the size
of the solution does not determines its stability as suggested in
\cite{b6}. Furthermore, as it can be seen from Fig. 6, our
oscillating solutions are such that the corresponding domain walls
becomes larger when one considers an increasing number of
oscillations between the regions of false and true vacua. Thus, if
one is analyzing the Lorentz invariant solutions, which evolve
from the false to the true vacuum, it can be concluded that the
time necessary for this transition is bigger for the higher
oscillating configurations. It would be very interesting to see if
such kind of richer structure is still present in the case of
brane world dominated scenarios \cite{prd2008}, similarly to what
happens with the cosmological instantons \cite{dunne}.

Before ending the work, we note that one can obtain a generalized
one-dimensional configuration having an arbitrary number of oscillating
regions, which can be written as

\begin{align}
& \phi (x)=S(-x+x_{1})(\epsilon -1)(1-e^{x-x_{1}})+S(x-x_{n-1})(\epsilon
+1)(1-e^{x_{n-1}-x})  \notag \\
& +\sum_{m=2}^{n-1}\left\{ S(x-x_{m-1})S(-x+x_{m})[\epsilon
+(-1)^{m}]\right.  \\
& \left. \times \lbrack 1-\frac{\sinh (x-x_{m-1})+\sinh (x_{m}-x)}{\sinh
(x_{m-}x_{m-1})}\right\} ,~n\geqslant 3.  \notag
\end{align}

In the above generalization of the kink-like solution, $S(x)$ is the
Heaviside function and $n$ stands for the number of regions. Now, for the
case of the lump (bubble) starting and finishing at the false vacuum, our
general solution is represented by

\begin{align}
\phi (x)& =S(-x+x_{1})(\epsilon -1)(1-e^{x-x_{1}})+S(x-x_{n-1})(\epsilon
-1)(1-e^{x_{n-1}-x})  \notag \\
& +\sum_{m=2}^{n-1}\left\{ S(x-x_{m-1})S(-x+x_{m})[\epsilon
+(-1)^{m}]\right.  \\
& \left. \times \lbrack 1-\frac{\sinh (x-x_{m-1})+\sinh (x_{m}-x)}{\sinh
(x_{m-}x_{m-1})}\right\} ,~n\geqslant 3.  \notag
\end{align}

Finally, the corresponding generalization for the configuration starting and
finishing at the true vacuum is given by

\begin{align}
& \phi (x)=S(-x+x_{1})(\epsilon -1)(1-e^{x-x_{1}})+S(x-x_{n-1})(\epsilon
+1)(1-e^{x_{n-1}-x})  \notag \\
& +\sum_{m=2}^{n-1}\left\{ S(x-x_{m-1})S(-x+x_{m})[\epsilon +(-1)^{m}]\right.
\notag \\
& \left. \times \lbrack 1-\frac{\sinh (x-x_{m-1})+\sinh (x_{m}-x)}{\sinh
(x_{m-}x_{m-1})}\right\} ,~n\geqslant 3.  \label{60.1}
\end{align}%
The solutions (26) and (\ref{60.1}) are represented in Fig. 6.

The two-dimensional case is written \ as

\begin{align}
\phi (r)& =S(r)S(-r+r_{1})(\epsilon +1)\left[ 1-\frac{I_{0}(r)}{I_{0}(r_{1})}%
\right]  \notag \\
& +S(r-r_{n-1})(\epsilon -1)\left[ 1-\frac{K_{0}(r)}{K_{0}(r_{n-1})}\right]
\notag \\
& +\sum_{m=2}^{n-1}S(r-r_{m-1})S(-r+r_{m})[\epsilon +(-1)^{m-1}]\left\{
(-1)^{m-1}\right.  \notag \\
& \left. +\frac{%
[K_{0}(r_{m-1})-K_{0}(r_{m})]I_{0}(r)+[I_{0}(r_{m})-I_{0}(r_{m-1})]K_{0}(r)}{%
K_{0}(r_{m})I_{0}(r_{m-1})-K_{0}(r_{m-1})I_{0}(r_{m})}\right\} ,~n\geqslant
3.  \label{61}
\end{align}

Finally, the case of three dimensions is such that

\begin{align}
\phi (r)& =S(r)S(-r+r_{1})(\epsilon +1)\left[ 1-\frac{r_{n}}{r}\frac{\sinh
(r)}{\sinh (r_{1})}\right]  \notag \\
& +S(r-r_{n-1})(1-\epsilon )\left( \frac{r_{n-1}e^{r_{n-1}-r}}{r}-1\right)
\notag \\
& +\sum_{m=2}^{n-1}\left\{ S(r-r_{m-1})S(-r+r_{m})[\epsilon +(-1)^{m-1}]
\left[ -\frac{(r_{m}e^{r_{m}}-r_{m-1}e^{r_{m-1}})}{(e^{2r_{m}}-e^{2r_{m-1}})}%
\frac{e^{r}}{r}\right. \right.  \notag \\
& \left. \left. +\frac{(r_{m}e^{r_{m-1}}-r_{m-1}e^{r_{m}})}{2re^{r}}\right]
\cosh (r_{m-r_{m-1}})+1\right\} ,~n\geqslant 3.  \label{62}
\end{align}

\textbf{Acknowledgements: }The authors thanks to CNPq and FAPESP for partial
financial support. ASD also give thanks to Professor D. Bazeia for
introducing him to the matter of solitons and BPS solutions. This work was
partially done during a visit (ASD) within the Associate Scheme of the Abdus
Salam ICTP.

\newpage

\begin{figure*}[tbp]
\centering
\includegraphics{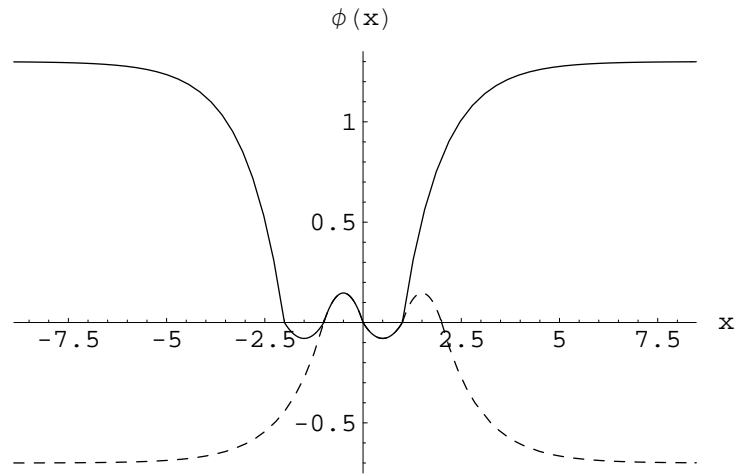}
\caption{Lumplike solutions (bubble) with two oscillations for $\protect%
\epsilon =0.3$.}
\label{fig7:bolhanossa1dfalsetruevacuum}
\end{figure*}

\begin{figure*}[tbp]
\centering
\includegraphics{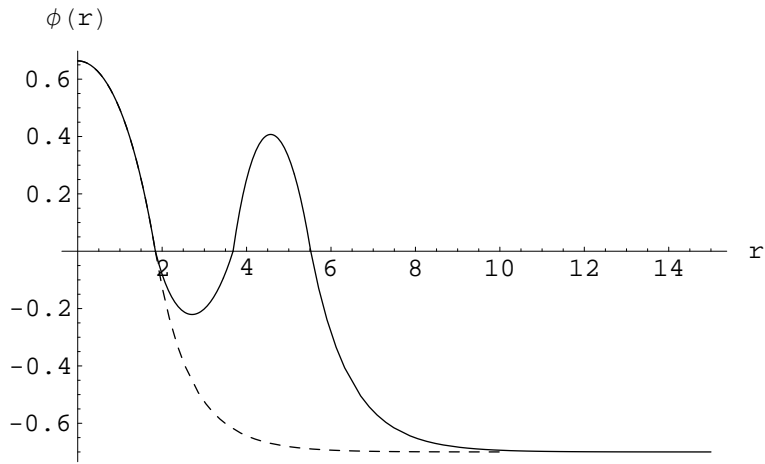}
\caption{Oscillating configuration in two dimensions for $\protect\epsilon %
=0.3$. The dashed line corresponds to the case studied in \protect\cite{b6}.}
\label{fig10:bol2d}
\end{figure*}

\begin{figure*}[tbp]
\centering
\includegraphics{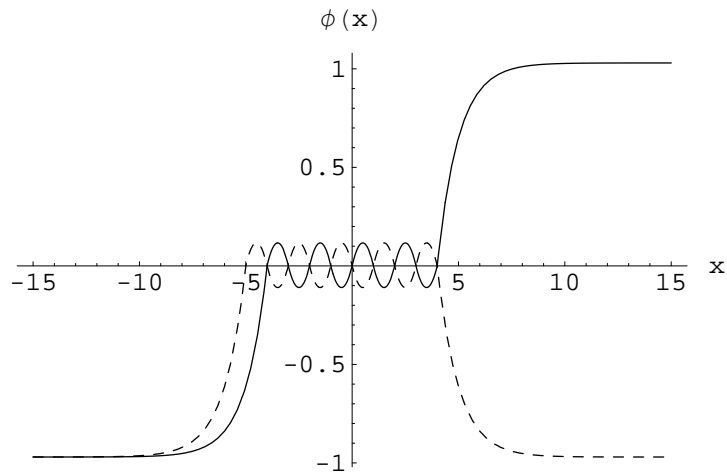}
\caption{Generalization of the kinklike (dashed line) and lumplike
(solid line) solutions with $\protect\epsilon =0.3.$}
\label{fig12:generalazaçãokinkbolha1D}
\end{figure*}

\begin{figure}[htbp]
\centering
\includegraphics{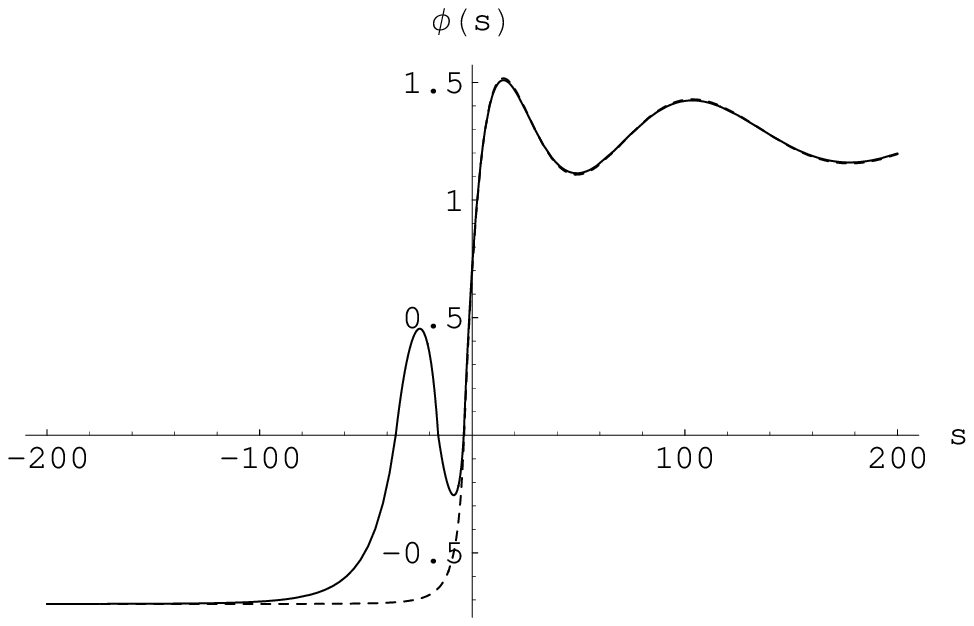} \label{fig:invlorentz}
\caption{Lorentz invariant solution for $\protect\epsilon %
=0.2828225347021907 $. The dashed line is the solution presented by
Theodorakis and the solid line is the one introduced here.}
\end{figure}


\begin{thebibliography}{99}
\bibitem{b1} G. B. Whitham, \textit{Linear and Non-Linear Waves}, John Wiley
and Sons, New York, (1974).

\bibitem{b2} A.C. Scott, F. Y. F. Chiu and D. W. Mclaughlin, Proc. I.E.E.E.
\textbf{61} (1973) 1443.

\bibitem{rajaramanweinberg} R. Rajaraman an E. J. Weinberg, Phys. Rev. D
\textbf{11} (1975) 2950.

\bibitem{b3} H. Arodz, Phys. Rev. D \textbf{52} (1995) 1082; Nucl. Phys.
\textbf{B450} (1995) 174; H. Arodz and A. L. Larsen, Phys. Rev. D \textbf{49}
(1994) 4154.

\bibitem{tetradis} A. Strumia \ and N. Tetradis, Nucl. Phys. \textbf{B542}
(1999) 719.

\bibitem{csaki} C. Csaki, J. Erlich, C. Grojean and T. J. Hollowood, Nucl.
Phys. B \textbf{584} (2000) 359.

\bibitem{gremm} M. Gremm, Phys. Lett. B \textbf{478} (2000) 434.

\bibitem{a1} A. de Souza Dutra and A. C. Amaro de Faria, Jr., Phys. Rev. D
\textbf{72} (2005) 087701.

\bibitem{shiffman} M. A. Shifman and M. B. Voloshin, Phys. Rev. D \textbf{57}
(1998) 2590.

\bibitem{bazeia1} D. Bazeia, M. J. dos Santos and R. F. Ribeiro, Phys. Lett.
A \textbf{208} (1995) 84. D. Bazeia, W. Freire, L. Losano and R. F. Ribeiro,
Mod. Phys. Lett. A \textbf{17} (2002) 1945.

\bibitem{campos} A. Campos, Phys. Rev. Lett. \textbf{88} (2002) 141602.

\bibitem{melfo} A. Melfo, N. Pantoja and A. Skirzewski, Phys. Rev. D \textbf{%
67} (2003) 105003.

\bibitem{PLB05} A. de Souza Dutra, Phys. Lett. B \textbf{626} (2005) 249.

\bibitem{PLB06} A. de Souza Dutra and A. C. Amaro de Faria Jr. , Phys. Lett.
B \textbf{642} (2006) 274.

\bibitem{bazeiaPLB06} V. I. Afonso, D. Bazeia and L. Losano, Phys. Lett. B
\textbf{634} (2006) 526.

\bibitem{giovannini} M. Giovannini, Phys. Rev. D \textbf{75} (2007) 064023;
Phys. Rev. D \textbf{74} (2006) 087505.

\bibitem{Rajaraman} R. Rajaraman, \textit{Solitons and Instantons}
(North-Holand, Amsterdam, 1982).

\bibitem{Vilenkin} A. Vilenkin and E. P. S. Shellard, \textit{Cosmic Strings
and Other Topological Defects} (Cambridge University, Cambridge, England,
1994).

\bibitem{cvetic} M. Cvetic and H. H. Soleng, Phys. Rep. \textbf{282} (1997)
159.

\bibitem{vachaspati} T. Vachaspati, \textit{Kinks and Domain Walls: An
Introduction to Classical and Quantum Solitons} (Cambridge University Press,
Cambrifge, England, 2006).

\bibitem{rajaraman} R. Rajaraman, Phys. Rev. Lett. \textbf{42} (1979) 200.

\bibitem{boya} L. J. Boya and J. Casahorran, Phys. Rev. A \textbf{39} (1989)
4298.

\bibitem{prd2008} A. de Souza Dutra, A. C. Amaro de Faria Jr. and M. Hott,
Phys. Rev. D \textbf{78} (2008) 043526.

\bibitem{b6.1} M. K. Prasad and C. M. Sommerfield, Phys. Rev. Lett. \textbf{%
35} (1975) 760; E. B. Bolgomol 'nyi, Sov. J. Nucl. Phys. \textbf{24} (1976)
449.

\bibitem{coleman} S. Coleman, Phys. Rev. D \textbf{15 }(1977) 2929.

\bibitem{coleman2} Curtis G. Callan, Jr. and S. Coleman, Phys. Rev. D
\textbf{16} (1977) 1762.

\bibitem{dunne} G. V. Dunne and Q. H. Wang, Phys. Rev. D \textbf{74} (2006)
024018.

\bibitem{horovitz} B. Horovitz, J. A. Krumhansl and E. Domany, Phys. Rev.
Lett. \textbf{38} (1977) 778.

\bibitem{deleo} S. E. Trullinger and R. M. DeLeonardis, Phys. Rev. A \textbf{%
20} (1979) 2225.

\bibitem{maki} H. Takayama and K. Maki, Phys. Rev. B \textbf{20} (1979) 5009.

\bibitem{b6} Stavros Theodorakis, Phys. Rev. D \textbf{60} (1999) 125004.

\bibitem{losano} D. Bazeia, A. S. In\'{a}cio and L. Losano, Int. J. Mod.
Phys. A \textbf{19} (2004) 575.

\bibitem{gompper} G. Gompper and S. Z. Zschocke, Phys. Rev. A \textbf{46}
(1992) 4836.

\bibitem{lohe} M. A. Lohe, Phys. Rev. D \textbf{20} (1979) 3120.

\end{thebibliography}
\end{document}